\documentstyle[aps,preprint,epsfig]{revtex}

\begin{document}
\draft
\def\be{\begin{equation}}
\def\ee{\end{equation}} 
\def\bfi{\begin{figure}}
\def\efi{\end{figure}}
\def\bea{\begin{eqnarray}}
\def\eea{\end{eqnarray}}
\title{Interface fluctuations, bulk fluctuations and dimensionality in the
off-equilibrium response of coarsening systems}

\author{Federico Corberi$^\dag$, Eugenio Lippiello$^\ddag$ and 
Marco Zannetti$^\S$}
\address{Istituto Nazionale per la Fisica della Materia,
Unit\`a di Salerno and Dipartimento di Fisica, Universit\`a di Salerno,
84081 Baronissi (Salerno), Italy}

\maketitle
 
\begin{abstract}

The relationship between statics and dynamics proposed by
Franz, Mezard, Parisi and Peliti (FMPP) for slowly relaxing
systems [Phys.Rev.Lett. {\bf 81}, 1758 (1998)]
is investigated in the framework of non disordered coarsening systems.
Separating the bulk from interface response we find that 
for statics to be retrievable from dynamics the interface
contribution must be asymptotically negligible.
How fast this happens depends on dimensionality. 
There exists a
critical dimensionality above which  
the interface response vanishes like the interface density  
and below which it vanishes more slowly.  
At $d=1$ the interface response does not vanish leading to
the violation of the FMPP scheme. 
This behavior is explained in terms of
the competition between curvature driven and field driven
interface motion.

\end{abstract}
 
 
Recently Franz, Mezard, Parisi and Peliti (FMPP) have 
proposed \cite{franz98,franz99} a connection
between static and dynamic properties of slowly relaxing systems. 
This is of 
fundamental importance for disordered systems such as spin glasses, 
since the low 
temperature equilibrium properties are hard to compute and still 
object of controversy 
after many years of intensive study. The existence of a bridge 
between statics and dynamics
then offers a useful alternative tool for the investigation of 
the equilibrium state.

Here we are interested in the analysis of this connection for 
non disordered coarsening systems, such as 
a ferromagnet quenched below the critical point. In this case the 
structure of the equilibrium
state is simple and well known. Therefore, these systems are particularly 
suitable for a 
detailed understanding of the method. Nonetheless, existing results are 
quite puzzling.
For the nearest neighbors Ising model numerical results 
in space dimension
$d\geq 2$ \cite{barrat98} 
fit into the FMPP scheme, while recent exact analytical 
results in the $d=1$
case \cite{godreche2000} show a qualitatively different behavior excluding any 
connection between
the relaxation properties and the structure of the equilibrium state. 
In this paper we investigate the problem through a 
careful analysis of the linear response function as the space dimensionality 
is changed. This allows to put together
the rich and interesting picture of what goes on in the off equilibrium 
response of a coarsening 
system and to uncover the mechanism whereby the FMPP scheme is or is not 
verified.
 
Let us first outline the problem. We consider a phase ordering system with 
a scalar order parameter quenched at the time $t=0$ 
from high temperature to a final temperature $T$ below the critical point. 
The time
evolution takes place by formation and coarsening of domains of the 
opposite equilibrium
phases \cite{bray94}. 
The characteristic feature of the process is the coexistence of 
fast and slow dynamics.
Within domains local equilibrium is reached rapidly while the coarsening 
process is slow. 
In the case of non conserved order parameter, as it will be considered here, 
the typical domain size 
grows like $L(t)\sim t^{1/2}$.
This phenomenology suggests the split of the order parameter into the sum 
of two independent components \cite{mazenko88}
\begin{equation}
\phi(\vec x,t)=\psi(\vec x,t)+\sigma(\vec x,t) 
\label{split}
\end{equation}
where $\psi(\vec x,t)$ describes equilibrium thermal fluctuations within 
domains, the ordering component $\sigma(\vec x,t)$ takes values $ \pm m_T$ 
in the bulk of domains  with 
the change of sign occurring at the interfaces and 
$m_T$ is the equilibrium spontaneous magnetization.
The split~(\ref{split}) accounts well for the observed behavior
\cite{bouchaud97} of the autocorrelation
function $C(t,t')=\langle \phi (\vec x,t)\phi (\vec x,t')\rangle$ 
where $t \geq t'\geq 0$.
This quantity can be written as the sum of two terms
\begin{equation}
C(t,t')=C_{st}(t-t')+C_{ag}\left ( \frac{t}{t'}\right )
\label{c}
\end{equation}
representing, respectively, the stationary dynamics of thermal 
fluctuations $ \psi( \vec x,t)$ and the slow out 
of equilibrium dynamics of
$ \sigma ( \vec x,t)$ obeying an aging form. At equal times  
$C_{st}(t-t'=0)=m_0^2- m_T^2$,
where $m_0$ is the $T=0$ spontaneous magnetization, and 
$ C _{ag}(1)= m_T^2 $. Furthermore,
due to the wide separation of time scales, in the range of time over which 
$C_{st}$
decays to zero the aging contribution remains practically constant 
$C_{ag}(t/t')\simeq m_T^2$.

Suppose, next, that at the time $t_w >0$ a small 
random field with expectations 
$ \overline{h(\vec x)}=0$, $\overline{h(\vec x)h(\vec y)}= 
h_0^2\delta(\vec x-\vec y)$ is switched on.
Given the structure~(\ref{split}), the perturbation affects only
the ordering component $\sigma (\vec x,t)$, leaving thermal
fluctuations unaltered, exactly as it happens in the equilibrium
ordered phase under the action of an external field.
Therefore, each unperturbed configuration is mapped into a new one 
$\phi (\vec x,t)\to \phi _h(\vec x,t)=\psi (\vec x,t) +\sigma _h (\vec x,t)$
with a modified ordering component. In general, the perturbation will
modify both the bulk and the interface behavior of $\sigma (\vec x,t)$.
Accordingly, to linear order we may write    
\begin{eqnarray}
\sigma _h(\vec x,t)&=&\sigma (\vec x,t)+\int d\vec x'\chi _B(\vec x-\vec x',t,t_                              w)h(\vec x') \nonumber \\
                   &+& \int d\vec x'\chi _I(\vec x-\vec x',t,t_w)h(\vec x') 
\label{chiB}
\end{eqnarray}
where the bulk response function $\chi _B$, accounting for the change in the 
magnetization within domains, must be related to thermal fluctuations via the 
equilibrium fluctuation dissipation theorem (FDT)
\begin{eqnarray}
T\chi _B & & (\vec x-\vec x',t,t_w) = \nonumber \\
         & & C_{st}(\vec x-\vec x',t-t_w=0) - C_{st}(\vec x-\vec x',t-t_w). 
\label{FDT}
\end{eqnarray}

At this point one can already get a glimpse on what conditions
must be realized for statics and dynamics to be connected. Since the
bulk response function $\chi _B$ probes the equilibrium fluctuations
and that is where the information on the equilibrium state is stored,
for the FMPP scheme to work the interface contribution $\chi _I$
must disappear. Current belief is that indeed this is what happens
\cite{franz99,barrat98,berthier99}
assuming that $\chi _I$ goes like the
interface density $\rho _I(t)\sim L^{-1}(t)\sim t^{-1/2}$.
However, to a closer scrutiny things are not so straightforward and
the interface contribution turns out to have more structure than
hitherto believed. In particular, there is an unexpected dependence
on dimensionality which, in the end, accounts for the discrepancies
in the Ising model mentioned above.
Before entering the calculation of $\chi _I$, it is useful to complete 
the outline of the problem illustrating a few points.

i) {\it FMPP scheme.} Averaging over the external field and the noise the
on site total response function is given by
\begin{equation}
\chi(t,t_w)=\frac{1}{h_0^2V}\int d\vec x\overline{\langle\phi _h(\vec x,t)
\rangle h(\vec x)}=\chi _B(t,t_w)+ \chi _I(t,t_w).  
\label{chi}
\end{equation}
In the FMPP scheme one assumes that this quantity obeys the out of equilibrium
generalization of the FDT proposed by 
Cugliandolo and Kurchan~\cite{cugliandolo93}, namely 
that for large times $\chi (t,t_w)$ depends on the time arguments
through the autocorrelation function $\chi (C(t,t_w))$.
Furthermore, one assumes that 
$\lim_{t\to \infty}\chi (t,t_w) = \chi _{eq}$, where $\chi _{eq}$ is 
the equilibrium 
response function.
With these two hypothesis, the connection between statics and dynamics is 
derived in the form
\begin{equation}
D(q)=\tilde P(q)
\label{ddiq}
\end{equation}
where $D(q)=-T_F\left [\frac{d^2\chi(C)}{dC^2}\right ]_{C=q}$ is the 
dynamical quantity, while $ \tilde P(q) = \lim_{h\to 0}P_h(q)$
is the static one.
$P_h(q)$ is the overlap distribution in the perturbed Gibbs 
state. We recall that for a ferromagnetic system 
in the unperturbed Gibbs state below $T_C$ the overlap 
distribution is given by \cite{mezard87}
\begin{equation}
P(q)=\frac {1}{2}\delta (q-m_T^2)+ \frac {1}{2}\delta (q+m_T^2).
\label{pdiq}
\end{equation}

ii) {\it Validity of the FMPP scheme.} 
If in~(\ref{chi}) we take into account only the bulk contribution 
$\chi _B$, neglecting $\chi _I$, then the FMPP scheme is verified. 
In fact, using~(\ref{c}) 
and recalling that $C_{ag}(t/t')\simeq m_T^2$ in the range of times where 
$C_{st}(t-t')$ decays to zero, from~(\ref{FDT}) follows
\begin{equation}
T\chi(C) = \left \{ \begin{array}{ll} 
                   C(t,t)-C(t,t_w)   & \mbox{for $m_T^2<C<m_0^2$} \\
                   m_0^2-m_T^2 & \mbox{for $C\leq m_T^2$}.
                   \end{array}
                   \right.
\label{spezzata}
\end{equation}
Eq.~(\ref{ddiq}) then implies $\tilde P(q)=\delta (q-m_T^2)$. 
The presence of one $\delta$-function instead of two as 
in~(\ref{pdiq}) does not mean that in the limit $h\to 0$ symmetry is 
broken. Rather, it means that from a response due only to 
thermal fluctuations it is impossible to distinguish one pure state from 
the other and everything goes as if symmetry was broken.
Numerical simulations of the Ising model for $d\geq 2$ \cite{barrat98}
show evidence for convergence toward the structure~(\ref{spezzata}) 
in the parametric plot of $\chi$ versus $C$ as $t_w$ becomes large (Fig.~1).
This indicates that the interface term 
$\chi _I$ in~(\ref{chi}) must be asymptotically negligible. 

iii) {\it Violation of the FMPP scheme.}
From the exact analytical solution of the $d=1$ 
Ising model one finds a  
different behavior. In $d=1$ it is important to realize 
that in order to make compatible the existence of a linear response regime, 
which requires $\frac {h_0}{T} <<1$ and therefore $T>0$, with the existence 
of an equilibrium state of the form~(\ref{pdiq}), one must take the limit
$J\to \infty$ in the ferromagnetic coupling. This in turn implies 
$m_T^2 = m_0^2$. Hence, if ~(\ref{spezzata}) were to hold, 
$T\chi(t,t_w)$ would vanish. Instead, one 
finds \cite{godreche2000}
$T\chi(C)=(\sqrt 2/\pi)\arctan \left [ \sqrt 2 \cot 
\left (\pi C/2 \right ) \right ]$
which yields
$D(q)=\pi \cos \left (\pi q/2 \right )
\sin  \left ( \pi q/2 \right )/
\left [2-\sin  \left (\frac{\pi}{2}q\right )\right ]^2$
and which is in no way related to~(\ref{pdiq}), as shown in Fig.~1.
Furthermore, notice that $J=\infty $ leads to the suppression
of thermal fluctuations within domains. 
Therefore, from the above analytical form and
$\chi (t,t_w)=\chi _I(t,t_w)$ follows
$\lim _{t\to \infty}T\chi _I(t,t_w)=\lim _{C\to 0} T\chi (C)=
1/\sqrt 2$. Therefore, 
the violation of the FMPP scheme in $d=1$ 
is due to an asymptotically dominant interface contribution.

Having made clear the necessity of investigating 
the relative importance of the bulk and interface 
terms as dimensionality is varied, we now introduce a 
semiphenomenological model for $\chi _I(t,t_w)$. 
This is based on the standard methods 
of the late stage theory in phase ordering kinetics \cite{bray94}, 
when only interface motion is of interest. 
Dropping the bulk term in~(\ref{chiB}) and defining      
$\sigma _I(\vec x,t)=\sigma(\vec x,t) + 
\int d\vec x'\chi _I(\vec x-\vec x',t,t_w)h(\vec x')$ 
we resort for this quantity to the time dependent Ginzburg-Landau
model without thermal noise
\begin{equation}  
\frac{\partial \sigma _I}{\partial t}= \nabla ^2\sigma _I+m_T^2\sigma _I-
\sigma _I^3+h(\vec x).
\label{lang}
\end{equation}
Next, in order to allow for the action of the field on the interface
motion, while keeping the domain saturation fixed at the unperturbed
level $\pm m_T$, we make the ansatz
\begin{equation}
\sigma _I(\vec x,t)=\frac{u(\vec x,t)}{\sqrt{1+\frac {u^2(\vec x,t)}{m_T^2}}}
\label{gaf}
\end{equation}
and we make an approximation 
of the gaussian auxiliary field (GAF) type \cite{bray94,bray93}.
The idea is that the non-linearity of the transformation~(\ref{gaf})
is enough to take care of the non-linearity in the problem and the
auxiliary field $u(\vec x,t)$ can be treated in mean field theory.
Inserting~(\ref{gaf}) in~(\ref{lang}) and linearizing the equation of
motion for $u(\vec x,t)$ we find 
\begin{equation}    
\frac{\partial u}{\partial t}= 
\nabla ^2u+m_T^2u-\frac{3}{m_T^2}
\langle (\nabla u)^2\rangle +h(\vec x)
\label{gaf2}
\end{equation}
where $\langle (\nabla u)^2\rangle$ is evaluated self-consistently. 
The key point is that the linear equation~(\ref{gaf2}) allows $u(\vec x,t)$
to grow unboundedly yielding via~(\ref{gaf}) $\sigma _I (\vec x,t)\simeq
m_T {\mbox sign} [u(\vec x,t)]$ which enforces 
the saturation of domains at the 
required unperturbed value. Making a further mean field approximation
through the replacement of~(\ref{gaf}) by
$\sigma _I(\vec x,t)=m_T\frac{u(\vec x,t)}{\sqrt {S(t)}}$
with $S(t)=\langle  u^2(\vec x,t) \rangle$,
we have $\chi _I(t,t_w)=m_T\chi _u(t,t_w)/\sqrt{S(t)}$ where $\chi _u$ 
is the response of the auxiliary field. 
Computing $\chi _u(t,t_w)$ and $S(t)$ 
from~(\ref{gaf2}) and defining $\chi _{eff}(t,t_w)$ by
$\chi _I(t,t_w)=\rho _I(t) \chi _{eff}(t,t_w)$
we find our main result
\begin{equation}
\chi _{eff}(t,t_w)=t^{1-\frac{d}{2}}F(t_w/t)
\label{gaf4b}
\end{equation}
with $F(x)=A\int _x^1dyy^{-\frac{d+2}{4}}(1-y+\frac{t_0}{t})^{-d/2}$. 
Here $A$ is a dimensionality dependent constant and 
$t_0$ is a microscopic time 
related to the momentum cutoff $\Lambda$ by $t_0=\Lambda^{-2}$. 
The meaning of Eq.~(\ref{gaf4b}) can be made transparent 
regarding the response of the system as due to a set of 
interfaces each contributing through an effective response 
$\chi _{eff}$.
From~(\ref{gaf4b}) follows that this quantity obeys
\begin{equation}
\chi _{eff}(t,t_w)\sim (t-t_w)^\alpha
\label{power}
\end{equation}
with $\alpha =0$ for $d>2$, $\alpha=1-d/2$ for $d<2$ and
$\chi _{eff}\sim \log (t-t_w)$ for $d=2$ \cite{nota1}. 
Eq.~(\ref{power}) applies both in the short ($t-t_w<<t_w$)
and in the large ($t-t_w>>t_w$) time region, with
a change in the prefactor taking place about $t-t_w\simeq t_w$.
The full time dependence of $\chi _{eff}(t,t_w)$ 
obtained by plotting out Eq.~(\ref{gaf4b}) for different values
of $d$ is displayed in Fig.~2a.
A completely analogous behavior is obtained in the Ising model.
We have computed $\chi _I(t,t_w)$ for $d=1,2,3,4$ in the Ising case by
suppressing spin flips in the bulk of domains and we have plotted
$\chi _{eff} (t,t_w)$ in Fig.~2b \cite{nota2} . 
The common features of Fig.~2a and
Fig.~2b may be summarized stating that in both cases $\chi _{eff}$
obeys the power law~(\ref{power}) and that there exists a critical 
dimensionality $d_c$ such that the exponent $\alpha $ is
zero for $d>d_c$ while it grows positive with decreasing
dimensionality for $d<d_c$, reaching in both cases the final value
$\alpha =1/2$ at $d=1$. The differences are that $d_c=2$ in the GAF
approximation, while from the Ising simulations 
there is a good evidence (Fig.~2b) for $d_c=3$.
One particular consequence
of this, on which we comment below, is that while in the former case
we have $\alpha =0$ with logarithmic growth at $d=2$, instead in the
$d=2$ Ising model one finds $\alpha =1/4$ (Fig.~2b). 

The interpretation of the above results goes as follows.
In the perturbed system 
interface motion is driven not only by curvature, 
as in the unperturbed case \cite{bray94},
but also by the external field. These two mechanisms compete. 
For $d>d_c$ the curvature mechanism dominates. 
The external field then affects only the 
spins strictly belonging to the interface.
Within a microscopic
time $\chi _{eff}$ saturates (see lines for $d=3$ in Fig.~2a and
for $d=4$ in Fig.~2b)
leaving the overall response $\chi _I$ to decrease as $\rho _I(t)$.
Instead, if dimensionality is low enough ($d\leq d_c$) to weaken the 
curvature mechanism
to the point that the field driven motion may start to play a role
in the response, then the single interface
response $\chi _{eff}$ grows with time like $t^\alpha$ counteracting 
the default decrease in $\chi _I$ 
due to $\rho _I(t)$. This effect can be understood realizing that when
interfaces are field driven the field 
produces a large scale optimization of domain positions with
respect to field configurations \cite{nota3}.
$d=1$ is the extreme case
where there is no more curvature mechanism. Then, interface
motion is entirely field driven yielding
$\chi _{eff}$ which grows like $t^{1/2}$ and compensates exactly
for $\rho _I(t)$ producing a non vanishing $\lim _{t\to \infty}
\chi _I (t,t_w)$.

In summary, we have investigated the relationship 
between the out of 
equilibrium response function and the structure of the equilibrium 
state for coarsening systems.
For the FMPP scheme
to hold, the out of equilibrium response must be dominated
by equilibrium fluctuations in the bulk of domains.
This we have made precise by separating the bulk from interface
response and by analyzing the behavior of the interface contribution
$\chi _I(t,t_w)$. 
On the basis of a GAF model we have found that there 
exists a critical dimensionality $d_c=2$ 
such that for $d>d_c$ $\chi _I (t,t_w)$ behaves 
like the interface density $\rho _I(t)$, 
while for $d<d_c$ it vanishes slower than $\rho _I(t)$ 
and does not vanish anymore
at $d=1$, yielding the violation of the FMPP scheme.
We have explained this behavior identifying $d_c$ as the dimensionality 
below which the external field competes effectively with
the curvature in driving interface motion.
The overall picture is confirmed by numerical results for
the Ising model, apart from the upward shift from $d_c=2$ to $d_c=3$ in
the critical dimensionality.
This means that in the Ising case the field driven mechanism competes
with the curvature even more efficiently than in the
GAF approximation. In particular, the field driven mechanism is
clearly observable in the $d=2$ Ising model where $\alpha =1/4$
makes the interface response much less preasymptotic than what
was estimated on the basis of the interface density argument.
Finally, the counterpart of the statement that a
persistent interface contribution leads to the violation of
the FMPP scheme, is that even if $\chi _I (t,t_w)$ vanishes asymptotically,
due to the field driven interface motion it may vanish so slowly
to hide the realization of the FMPP scheme.

{\bf Acknowledgments}

We acknowledge useful discussions with A. Coniglio, U.
Marini Bettolo and L. Peliti. This work was partially
supported by the European TMR Network-Fractals Contract
No. FMRXCT980183. F.C. acknowledges support by INFM PRA-HOP 1999.

\dag corberi@na.infn.it \ddag lippiello@sa.infn.it 

\S zannetti@na.infn.it

\begin{figure}
\centerline{\psfig{figure=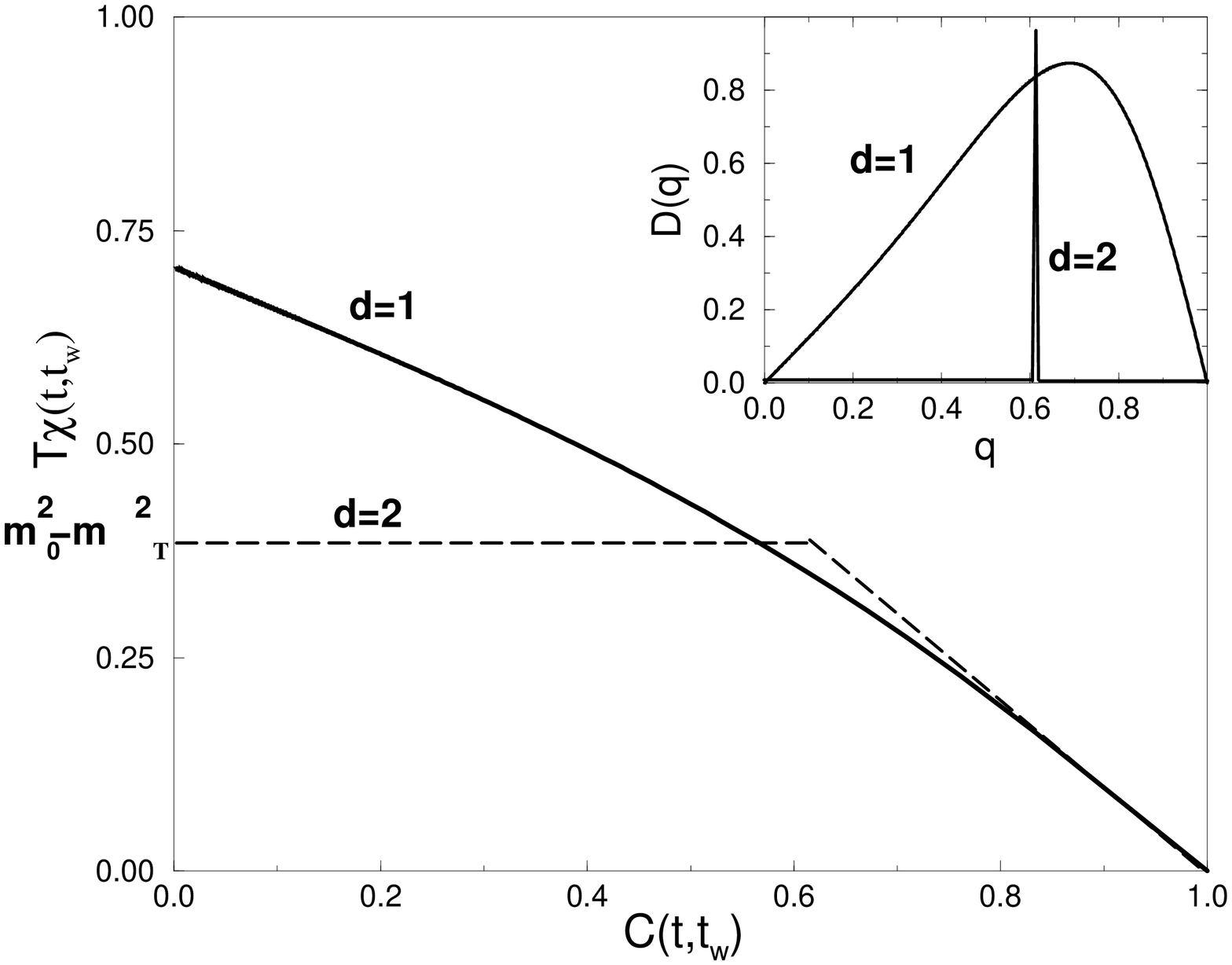,width=9cm,angle=-90}}
\caption{$T\chi (t,t_w)$ for the Ising model. 
For $d=2$ the curve shows the expected
behavior on the basis of Eq.~(\ref{spezzata}) 
for $t_w\to \infty $ with $T=2.2, J=1$.
For $d=1$ the curve is the plot of 
the analytic form of Ref.~[4].
The inset shows the corresponding $D(q)$.}
\label{Letter_fig1}
\end{figure}

\begin{figure}
\centerline{\psfig{figure=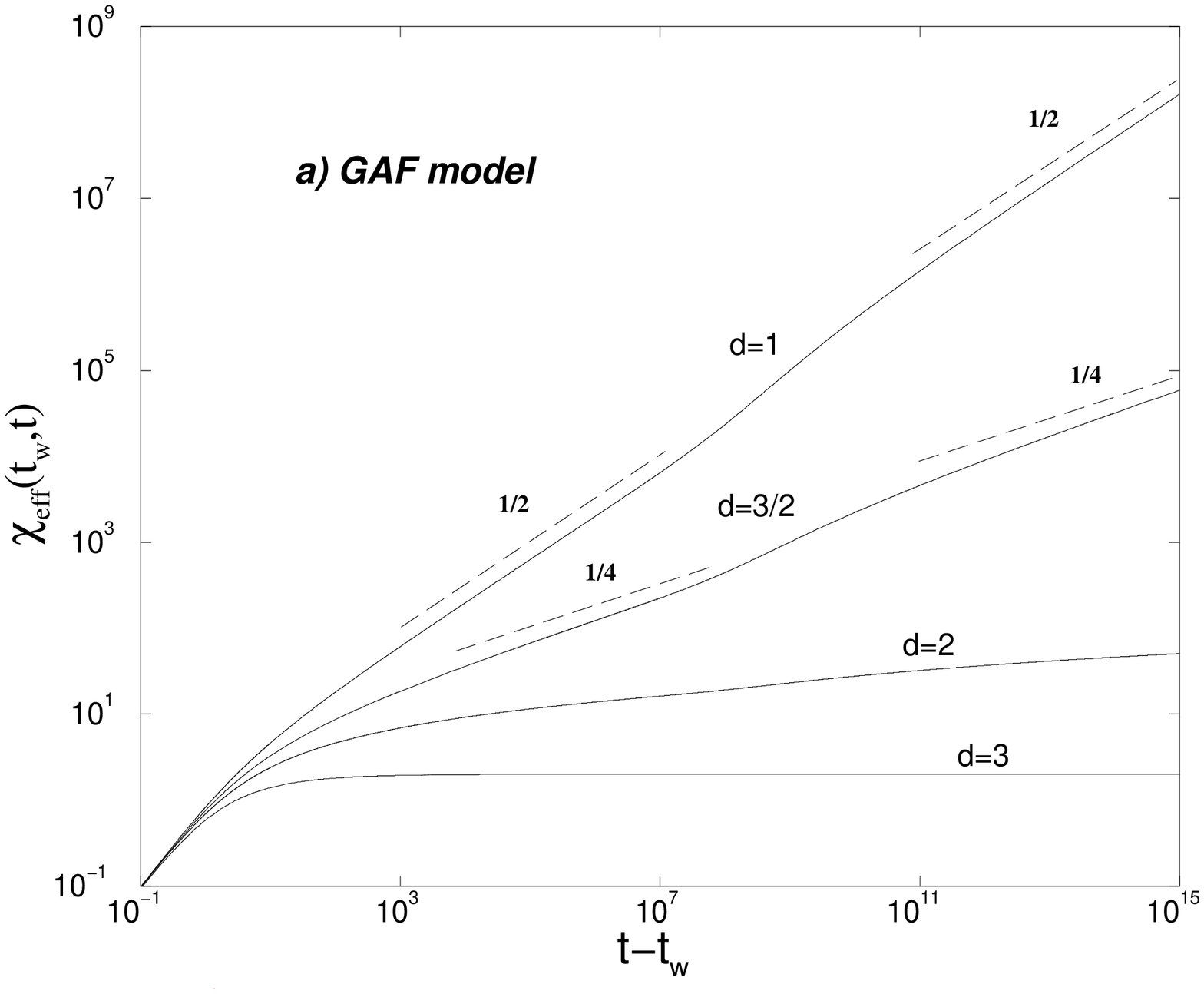,width=9cm,angle=-90}}
\centerline{\psfig{figure=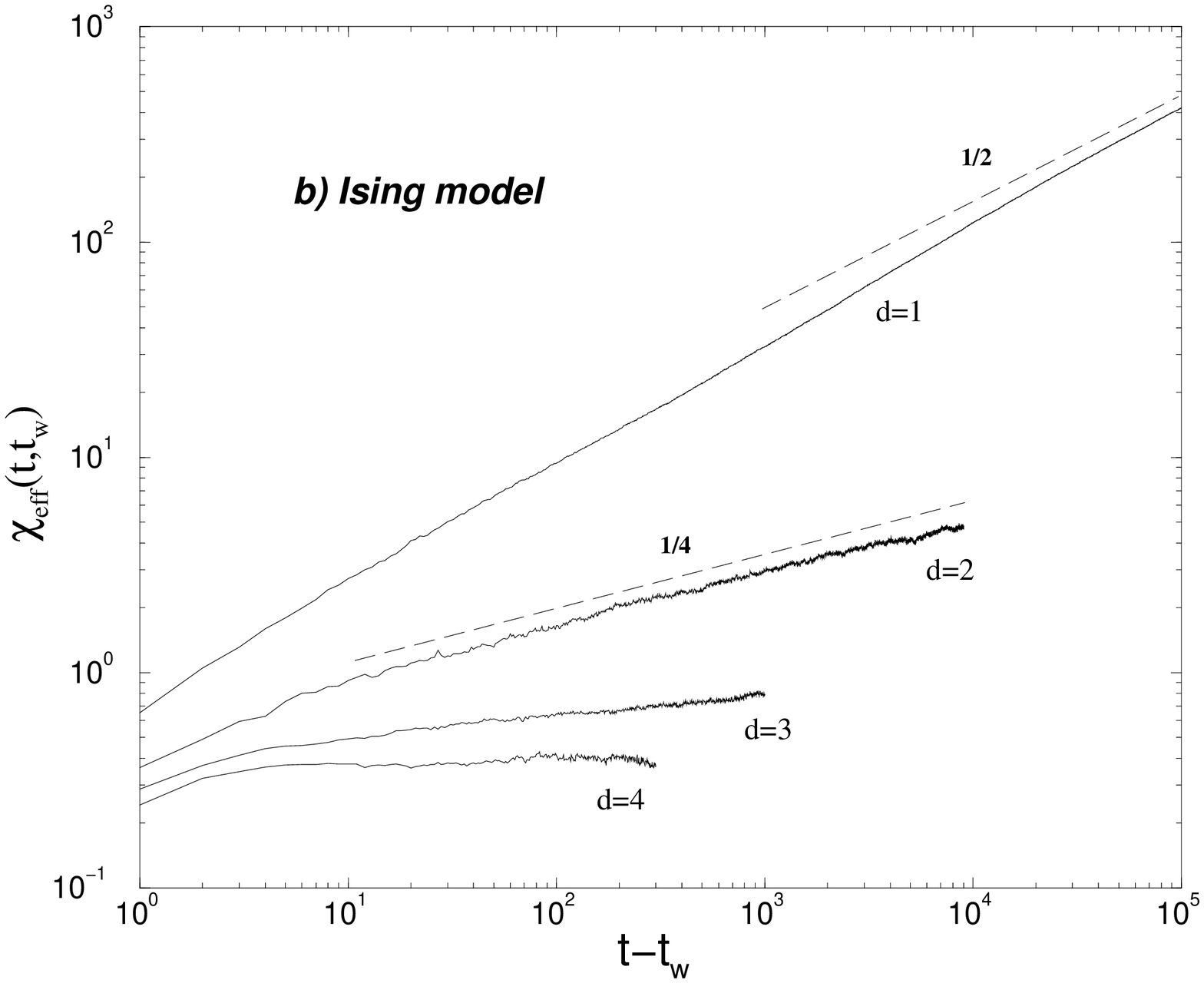,width=9cm,angle=-90}}
\caption{$\chi _{eff}(t,t_w)$ in the semiphenomenological
model with $t_w=10^8$ (a) and in the Ising
model without spin flips in the bulk (b).
For $d=1,2,3,4$, the temperature, waiting time and linear system sizes $L$ 
of the simulations are 
$T=0.48, 2.2,3.3,4.4$, $t_w=10^3,10^3,10^2,10$ and
$L=10^6,512,128,42$ with $J=1$ and averages 
over $170,6045,2590,922$ realizations.
The dashed lines are power laws 
with the corresponding exponent $\alpha $.
For $d=3$ in (b) the curve is very well fitted by $0.33+0.066\log (t-t_w)$.}
\label{newFig1}
\end{figure}


\begin{thebibliography}{99}
\bibitem{nota1}
A logarithmic correction in the response function for $d=2$ 
was first derived in Ref.~\cite{berthier99}.
\bibitem{nota3}
This can be checked very accurately by computing the response function
in simulations without bulk flip and with initial configurations
containing only one interface.
\end{thebibliography}
\end{document}